**Thermodynamic calculations using reverse Monte Carlo: A computational workflow for accelerated construction of phase diagrams for metal hydrides**

Swati Rana[1], Dayadeep S. Monder[1] and Abhijit Chatterjee[2]*

[1]Department of Energy Science and Engineering, Indian Institute of Technology Bombay, Mumbai, India, 400076

[2]Department of Chemical Engineering, Indian Institute of Technology Bombay, Mumbai, India, 400076

Email: abhijit@che.iitb.ac.in

**Abstract**

Metal hydrides are promising candidates for hydrogen storage applications. From a materials discovery perspective, an accurate, efficient computational workflow is urgently required that can rapidly analyze/predict thermodynamic properties of these materials. The authors have recently introduced a thermodynamic property calculation framework based on the lattice reverse Monte Carlo (RMC) method. Here the approach is extended to metal hydrides, which exhibit significant volume expansion, strong interaction between hydrogen and the host atoms, lattice strain, and a phase transition. We apply the technique to the nickel hydride ($NiH_x$) system by calculating the pressure-composition-temperature (PCT) isotherm and constructing its phase diagram. An attractive feature of our approach is that the entire phase diagram can be accurately constructed in few minutes by considering <10 configurations. In contrast, a popular technique based on grand canonical Monte Carlo would require sampling of several million configurations. The computational workflow presented paves the way for the approach to be used in future for wider materials search and discovery.

1. Introduction

Reversible hydrogen storage is critical for realization of hydrogen powered technology [1–6]. Since the last decade, the search for hydrogen storage materials that can meet the technical targets set by the U.S. Department of Energy (DOE) for onboard storage applications has intensified [7,8]. Due to their large hydrogen storage capacity, metal hydrides (Pd, Mg, Al, Li, Ni, and their alloys) have emerged as attractive options [9–11]. Certain enhancements in thermodynamic properties of existing materials are being explored [12–19], e.g., alloying with catalysts and nanosizing [20–31], to increase the capacity and to lower the operating temperature. However, the discovery of new materials is hindered by the large search space comprising of multiple target properties (hydrogen storage capacity, uptake/release, cost, cycle life, etc.) and factors affecting these properties [15,32]. It is hoped that computational techniques will play a key role [33–35] in speeding-up the discovery and optimization of these materials.

As far as thermodynamic properties are concerned, while the free energy associated with ideal gas, translational, rotational and vibrational degrees of freedom can be evaluated reasonably easily [36], computing the configurational term for the pressure-composition-temperature (PCT) isotherm and phase diagram is a challenge. Diverse hydrogen arrangements are possible in these disordered solid phases. Typically, several millions of configurations need to be sampled using standard grand canonical Monte Carlo (GCMC) [37,38] before the equilibrium configuration is obtained. This makes GCMC prohibitively expensive computationally for systems with >10,000 atoms because of the low acceptance probability of hydrogen insertion. Methods based on mean-field assumption possess a much lower computational overhead but these are not regarded as accurate enough for systems like metal hydrides that exhibit phase transitions [36,39]. Therefore, fast computer simulation methods are urgently needed that can correctly incorporate configurational terms like GCMC but possess a negligible computational overhead. We address this gap by exploring the application of a novel reverse Monte Carlo (RMC) based thermodynamics property calculation framework [40–43] to hydrogen storage materials. Using bulk Ni hydride as a prototype system, we present and discuss a computational workflow, which can be used to construct phase diagrams in the matter of few minutes.

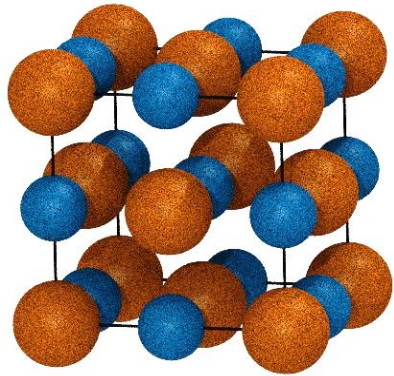

Figure 1. Nickel hydride crystal structure. $Ni$ atoms are shown in orange. Hydrogen sites (blue) can be either filled or vacant.

A non-stochiometric interstitial hydride ($NiH_x$) is formed by incorporating $H$ at the octahedral sites [44] (Figure 1). $x$ denotes the fraction of sites occupied by $H$. The $\alpha$-phase is obtained at low $H_2$ pressures – it contains a small amount of hydrogen at the interstitial sites. The hydrogen-rich ($x > 0.8$) $\beta$-nickel hydride phase forms at >600 MPa at 298K [45]. The $\beta$-phase crystallizes in cubic NaCl structure. The lattice constant of $Ni$ and $NiH$ is 3.53 and 3.731 Å, respectively. While constructing the PCT isotherm one needs to consider the lattice expansion.

Traditionally, RMC simulations have been employed in structural modelling [46–48], i.e., for generating 2D/3D atomistic structures with given short-range order (SRO) parameters. SRO parameters, such as, pair probabilities and radial distribution function, provide statistical information about the local atomic arrangement [38,42,49,50]. Recently, the authors have extended the RMC approach to the calculation of SRO parameters at equilibrium for specified composition, interaction and temperature [43]. Once the equilibrium configuration can be generated, quantities like chemical potential, free energy and phase diagrams are readily evaluated. Excess properties are correctly captured since the complete atomistic details are available [51]. It should be noted that previous demonstrations of the method [40–43] have been restricted to lattice systems with only first nearest neighbor (1NN) pair interactions being present and the material system does not undergo any volume change. The hydriding process in $Ni$ involves a significant volume change of 18%. Moreover, the interactions between $Ni$ and $H$ are modelled in this work using an embedded atom method (EAM) potential where the interactions are many-body in nature and extend to beyond 1NN. Therefore, application of the RMC method to the nickel hydride system is interesting as it also extends the method to a broader range of materials problems.

In this work, the central philosophy is as follows. RMC configurations can be created for given $x$ and SRO parameters $z$. The equilibrium configuration is determined by evaluating a quantity called the SRO growth rate $g(x, z, w, T)$. By specifying the hydrogen concentration $x$, system interactions $w$ and temperature $T$, the SRO parameter $z = z_{eq}$ that is associated with zero growth rate is determined. $z_{eq}$ provides the equilibrium configuration [41,52]. For the system size considered here (between 0.1-0.2 million atoms), typically several billions of configurations need to be sampled in GCMC before equilibrium is reach. The inherently sequential sampling procedure renders GCMC calculations expensive. In some cases, GCMC may not even properly converge due to geometric frustration [53]. In contrast, we show that using RMC the equilibrium configuration can be determined by probing four configurations or fewer. This makes the RMC approach particularly attractive for the construction of phase diagrams.

The outline for this paper is as follows. Section 2 discusses the theory of the RMC-based approach. The computational details are provided in Section 3. The results are discussed in Section 4 and Section 5 lists the main conclusions.

## 2. Theory

Ignoring for a moment the background Ni atoms and considering only the octahedral sites, nickel hydride can be treated as a binary system denoted as $H_xV_{1-x}$, where $V$ is a vacant site. In our simulations, the number of $Ni$ atoms $N_{Ni}$ is kept fixed, $N_H$ and $N_V$, the number of occupied and vacant sites, respectively, are specified, such that $x = N_H/N_t$ and $N_H + N_V = N_t$. Also, $N_t (= N_{Ni})$ refers to the number of octahedral sites. The hydrogen atom chemical potential in gas phase $\mu_H$ at temperature $T$ that is in equilibrium with such a system is determined for the given $x$ by considering the configurational (or equilibrium SRO) and vibrational terms. This contrasts with a GCMC calculation where the $\mu_H$ is specified and $x$ is determined. Interactions between $Ni$ and $H$ are described using an embedded atom method (EAM) potential developed by Angelo et al. [54].

### 2.1 Short-range order parameter

On-lattice nickel hydride structures can be constructed using RMC [43] by specifying the first nearest neighbor (1NN) pair probability $z$, i.e., the probability of finding another $H$ at position of an $H$ atom. The target number of 1NN $H - H$ pairs in the RMC configuration is

$$N_{HH} = \frac{1}{2}cN_txz. \qquad (1)$$

$z$ is an SRO parameter. $c$ is the coordination number. For $NiH$, $c = 12$. From equation (1), it is clear that $x$ and $z$ dictate the 1NN pair count. An important assumption made is that $z$ is a descriptor for the local $H$ arrangement. This implies that the converged RMC configuration is assumed to provide a reasonable estimate of the pair counts for 2NN, 3NN positions and so on, as well as various triplet counts and other many-atom cluster arrangements. Therefore, these quantities are also a function of $x$ and $z$. A test to ascertain whether this assumption is correct was discussed in Ref. [41]. We shall discuss these points later.

To generate a 3D configuration in RMC, a lattice with $N_t$ sites, the fractional occupation $x$ and the SRO parameter $z$ are provided. The initial guess $H$ and $V$ arrangement is a perfectly random, well-mixed one. A sequence of trial swap moves is performed, such that the positions of a randomly selected pair of $H$ and $V$ are interchanged with a certain acceptance probability, until the target number $N_{HH}$ is achieved (Section 3.1). The swap moves result in tuning of the SRO parameter till the desired $z$ is reached. RMC calculations are independent of temperature and atomic interactions, and the time required on a desktop computer for convergence is typically in seconds for the system size (0.1-0.2 million atoms) considered here [43]. If required, off-lattice structures can be generated by performing a short NPT-molecular dynamics (constant pressure and temperature MD) followed by energy minimization, which relaxes the structure.

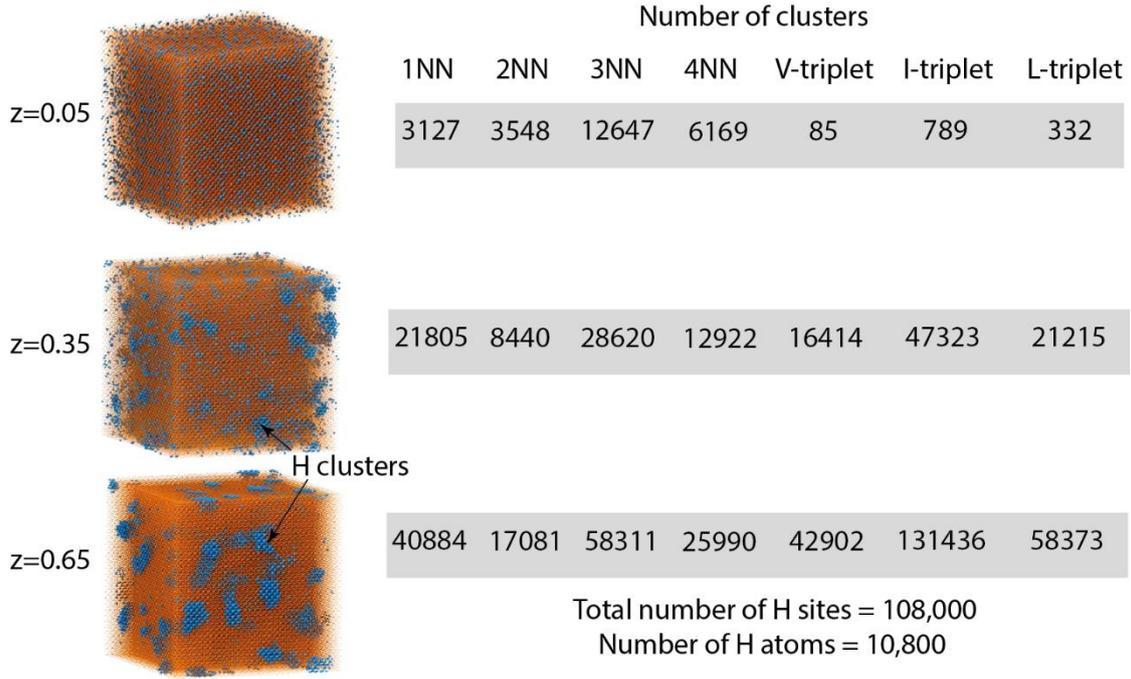

figure 2. RMC configurations generated by specifying $N_t$, $x$ and $z$. Hydrogen distribution for $x = 0.1$ (orange: Ni and blue: H atoms) with SRO parameter being 0.05, 0.35 and 0.65. Higher $z$ result in clustering of H atoms. Definition of clusters are provided in Section 4.1.

Figure 2 shows a few examples of RMC configurations. Since RMC employs random seeds, the configurations generated are random. However, the probability distribution of local arrangements are deterministic functions of $x$ and $z$ [43]. For a well-mixed arrangement, $z = x$. Such a situation is encountered when interactions between $H$ atoms are absent. Attractive interactions will cause the $H$ atoms to cluster, in which case, $z > x$. From Figure 2, it is evident that an RMC configuration for a given $x$ and $z$ will also provide the number of pairs, triplets and other many-atom clusters as well as certain configurational properties, such as the energy. RMC configurations are used for construction of cluster expansion models (Section 4.1) and for calculation of lattice parameter and vibrational frequencies. Thus, the material properties are also functions of $x$ and $z$.

## 2.2 SRO growth rate associated with a configuration

A procedure for determining the equilibrium SRO parameter $z_{eq}$ is discussed in this section. The goal is to identify $z_{eq}$ for the given composition, interactions and temperature, i.e., $z_{eq} \equiv z_{eq}(x, \{w\}, T)$. Here $\{w\}$ denotes the complex many-body interactions in the $NiH_x$ system. This goal is achieved by following the procedure in Ref. [41].

The evolution of a configuration can be expressed in terms of the number of 1NN H-H pairs as:

$$\frac{dN_{HH}}{dt} = g(x, z). \qquad (2)$$

Here $t$ is a fictitious time and $g(x, z)$ is the SRO growth rate. Since $z$ is related to $N_{HH}$, an equivalent form of Equation (2) can be written where the left-hand side is $dz/dt$. Thus, Equation (2) is a probabilistic model. For our purpose, the evolution of a specific configuration need not be solved. Rather, $z_{eq}$ is determined through an interpolation procedure where first, $g(x, z)$ is calculated at a handful of $z$ points. For instance, Figure 2 shows the configurations for three different $z$ values (0.05, 0.35 and 0.65) with $x = 0.1$, and the corresponding growth rate needs to be determined. Next, one identifies $z_{eq}$ numerically where

$$g(x, z_{eq}) = 0. \qquad (3)$$

It should be noted that since $z$ is a probability, its value should lie between 0 and 1. In practice, topological constraints introduced by the lattice causes $z$ to have a smaller range of values [42]. For this reason, it should be checked whether the number of 1NN H-H pairs matches the target $N_{HH}$ to ensure proper convergence of the RMC calculation.

To obtain an expression for the growth rate, we consider moves in the configuration where an $H$ atom is made to hop to an adjacent vacant site. The transition rate $r_{i \to j}$ for a move from site $i$ to site $j$ is written in a manner analogous to the standard Metropolis one:

$$r_{i \to j} = \min(1, \exp(-\beta \Delta U_{i \to j})), \quad (4)$$

which captures thermodynamic aspects. $\beta = (k_B T)^{-1}$, $k_B$ is the Boltzmann constant and $\Delta U_{i \to j}$ denotes the change in energy. These hop moves are not executed but are considered only for the calculation of the growth rate. $r_{i \to j}$ is zero if the move is not possible (site $i$ is vacant and/or site $j$ is filled).

The growth rate for a configuration is calculated as

$$g(x, z) = \sum_{i \to j} r_{i \to j} \nu_{i \to j}. \quad (5)$$

$\nu_{i \to j}$ is a stoichiometric coefficient that denotes the change in the number of 1NN pairs because of the move. All moves possible in the current configuration are considered in the sum. According to Equation (2), a positive value of the growth rate implies that $z$ will increase and the configuration will evolve such that more 1NN $H - H$ pairs are formed. For the zero-growth rate, the system does not evolve. From Equation (3)-(5), it becomes evident that $z_{eq}$ is dictated by $x, w$ and $T$. Also, $g \equiv g(x, z, \{w\}, T)$. Thus, the traditional ability of RMC to solve the inverse problem of finding a 3D configuration consistent with the given SRO parameters is augmented with the ability to determine $z_{eq}$ using Equations (3)-(5).

The large system size (>0.1 million atoms) ensures that the distribution for the configurational quantity $\Delta U_{i \to j}$ is properly sampled, i.e., finite size effects are missing. The calculation of $\Delta U_{i \to j}$ requires the energy minimized final and initial states, and this quantity needs to be evaluated for several hop moves. Fortunately, this problem can be easily parallelized on a computer due to its linear scaling behavior with respect to the number of computer processors.

In the present study, we employ a cluster expansion model (CEM) [39,55,56] to estimate $\Delta U_{i \to j}$ to help us better understand the nature of $H-H$ interactions (see Section 4.1). The CEM provides the configurational energy in terms of pair and triplet counts while capturing the effect of lattice expansion, i.e., $\{w\}$ is a function of the hydrogen content.

## 2.3 Free energy calculations and phase diagram construction

Once the RMC configuration for $z_{eq}$ is available, the thermodynamic properties are estimated in the canonical ensemble, i.e., $N_H, N_t$ and $T$ are held constant. For the PCT isotherm, the $\mu_H$ is evaluated using the particle insertion method (PIM) as [43]

$$\beta \mu_H = \ln \frac{x}{1-x} - \ln \langle \exp(-\beta \Delta U_{ins,s}) \rangle. \tag{6}$$

$\Delta U_{ins,s}$ is the energy of inserting a hydrogen atom at vacant site $s$. The angular bracket signifies the mean value over all vacant sites in an equilibrium configuration generated using RMC. The first term represents the ideal term whereas the second represents the excess term due to $H-H$, $H-Ni$ and $Ni-Ni$ interactions (or $\{w\}$ in Section 4.1). Equation (6) includes configurational aspects. The Gibbs free of mixing $\Delta_{mix} G$ (units of eV/atom) is evaluated using the lattice chemical potentials. A common tangent approach is used with the $\Delta_{mix} G(x)$ curves to determine the concentration of H in the coexisting $\alpha$ and $\beta$ as a function of $T$ (section 4.3). The vibrational contribution to the Helmholtz free energy $F_{vib}$ [57], namely

$$F_{vib} = \sum_{i=1}^{3N} k_B T \left\{ \frac{\hbar \omega_i}{2T} + \ln \left( 1 - e^{-\frac{\hbar \omega_i}{T}} \right) \right\}, \tag{7}$$

is considered separately. $\omega_i$ is the vibrational frequency. The vibrational chemical potential is calculated as

$$\mu_{vib} = \left(\frac{\partial F_{vib}}{\partial N_H}\right)_{N_t,V,T}. \tag{8}$$

These vibrational corrections are added to $\mu_H$.

The gas phase hydrogen pressure $p_{H_2}$ is calculated using the relationships $\mu_H = \frac{1}{2}\mu_{H_2}$ [36] and

$$\mu_{H_2}(p_{H_2}, T) = \mu_{H_2}^0 + k_B T \ln\left(\frac{p_{H_2}}{P^0}\right), \tag{9}$$

$$\mu_{H_2}^0 = -k_B T \ln\left[\frac{2\pi mkT}{h^2}\right]^{\frac{3}{2}} k_B T - k_B T \ln\left[\frac{8\pi^2 I k_B T}{\sigma h^2}\right] + k_B T \ln\left(1 - e^{\frac{-h\nu}{k_B T}}\right) \tag{10}$$

$$- k_B T \ln w_e - D_e + \frac{h\nu}{2}.$$

The dissociation energy of H₂ molecule $D_e = -4.56$ eV and the vibrational frequency $\nu = 4161 \, cm^{-1}$. $P^0$ is 1 atm, $m$ is mass of H₂, $h$ is the Planck constant, $k_B$ is the Boltzmann constant, $I$ is the moment of inertia for diatomic hydrogen, $\sigma$ is the symmetry number for hydrogen = 2.

## 3 Computational Details

### 3.1. RMC simulations

We use the RMC algorithm as described in Ref. [43]. As noted earlier, for a given value of SRO parameter $z$, the number of $H - H$ bonds expected is $N_{HH,target} = \frac{1}{2}N_H cz$. The steps in RMC are as follows:

Step 1: $N_{HH}$ is calculated for the current configuration. The distance from the target structure is $d_t^2 = |N_{HH} - N_{HH,target}|^2$.

Step 2: A pair of $H$ and $V$ sites are randomly selected, and their positions are swapped. The new number of $H - H$ bonds $N_{HH,new}$ as well as the distance to target structure $d_{t,new}^2$ is evaluated.

Step 3: The acceptance probability is defined as

$$p_{acc} = \min\left(1, \exp\left(d_t^2 - d_{t,new}^2\right)\right). \tag{11}$$

Step 4: The lattice structure is updated if the move is accepted. Otherwise, the old configuration remains. Steps 1-4 are repeated in the next iteration.

The RMC calculation is stopped once two million swap moves have been attempted after the point that the target structure is first reached i.e., $d_t$ has become zero. This ensures proper convergence of the RMC calculation [51]. A periodic lattice containing $30 \times 30 \times 30$ unit cells with $N_{Ni} = 108,000$ is employed in RMC. The total number of atoms for $NiH$ system is 216,000. The use of a large system ensures that finite-size effects and the sampling error are negligible [42].

### 3.2 Vibrational chemical potential

A host lattice containing $8 \times 8 \times 8$ unit cells with $N_{Ni} = 2048$ atoms is considered. Periodic boundary conditions are used. Several configurations in the range $x \in [0,1]$ are considered. The vibrational modes were obtained by diagonalizing the dynamical matrix of the energy minimized structure. All calculations are performed with EAM potential parameterized by Angelo et al. [54]. Subsequently, the vibrational chemical potential is obtained using equations (7) and (8).

### 3.3 Overview of the computational workflow

The overall computational workflow comprises of the following home-grown packages:

I. Generating RMC configurations for given $x$ and $z$.

II. Training a cluster expansion model (CEM) using principal component regression (Section 4.1). The CEM predicts the $NiH_x$ system energy in terms of the cluster counts in a given RMC configuration.

III. Calculating the SRO growth rate $g(x, z, \{w\}, T)$ and determining $z_{eq}(x, \{w\}, T)$.

IV. Determining the vibrational chemical potential.

V. Using the particle insertion method to calculate $\mu_H(x, \{w\}, T)$. Vibrational corrections are included.

VI. Determining the Gibbs free energy of mixing $\Delta_{mix} G$ and using common tangent method to construct the phase diagram.

These packages are written in Fortran 90/95.

## 4. Results

### 4.1 CEM construction with the aid of RMC

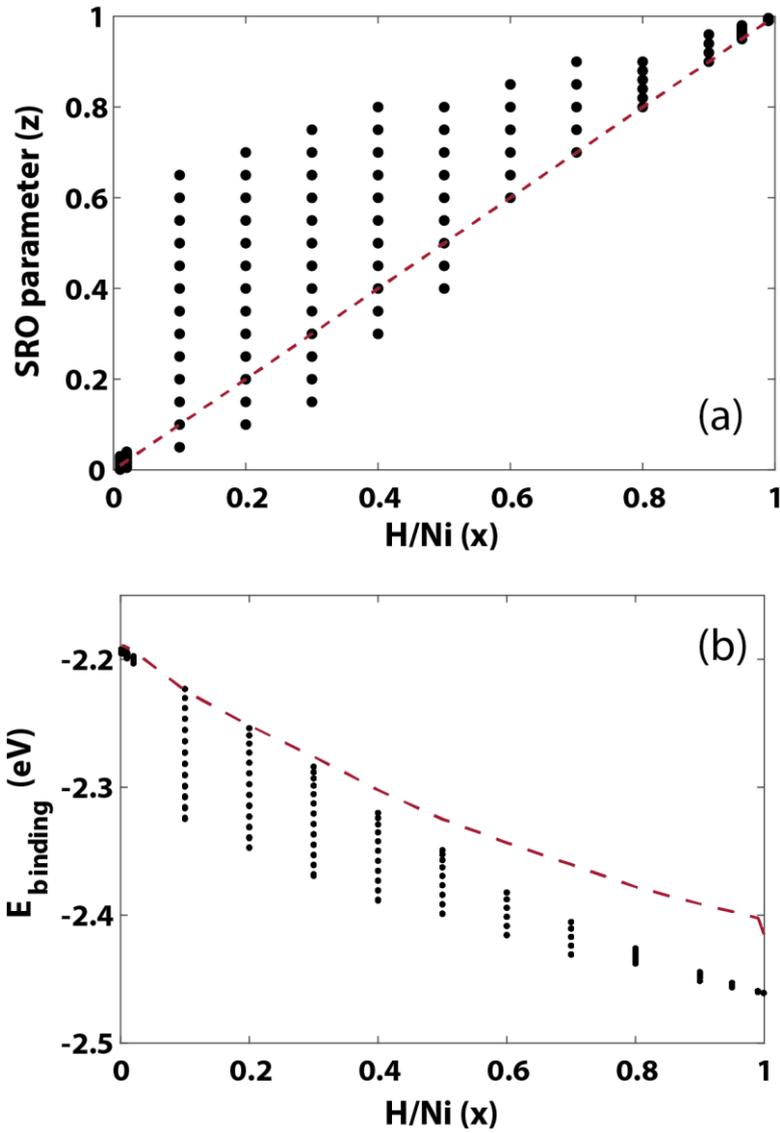

Figure 3. (a) Points $(x, z)$ used for construction of RMC configurations and subsequently for CEM database generation. (b) Binding energy ($E_{binding}$) per hydrogen atom as a function of $x$ obtained for different RMC configurations. Dashed line corresponds to the energy without considering solute-solute interactions, i.e., $E_{binding}^0(x)$.

The EAM potential developed by Angelo et al. [54] is used to build a cluster expansion model. Figure 3a shows points in the $x − z$ space sampled for constructing RMC configurations. 16 different hydrogen concentrations $x$ were sampled, $x \in [0.001, 0.999]$. Each configuration was relaxed and $E_{NiH_x}$ is computed. We define the binding energy per $H$ atom as:

$$E_{binding}(x,z) = \frac{E_{NiH_x}(x,z) - E_{Ni} - N_H E_H}{N_H}. \tag{12}$$

Here, $E_{NiH_x}$ is the energy of a relaxed off-lattice configuration, $E_{Ni}$ is the energy of the Ni bulk system and $E_H$ is the energy of a hydrogen atom in vacuum (taken to be 0 eV). $E_{binding}$ varies as a function of $x$ (Figure 3b). There are two main observations. First, the binding energy increases in magnitude from -2.19 to -2.45 eV/atom as more $H$ is added. This value of binding energy is reasonable: using the experimentally reported heat of solution for $H$ in $Ni$ ($\beta$-phase) of 12.5 kJ/mol [45] and dissociation energy of gas phase $H_2$ molecule ($D_e = -4.56$ eV) yields a binding energy of -2.41 eV/atom. Second, a significant spread in the binding energy is observed for a given $x$. This implies that a simple nonlinear function in terms of $x$ cannot provide a reasonable estimate for $E_{binding}$ and that the local $H$ arrangement needs to be considered. We write

$$E_{binding}(x,z) = E^0_{binding}(x) + \frac{1}{N_H} E_{many-body}(x,z). \tag{13}$$

$E^0_{binding}(x)$ captures the interaction of a single $H$ atom with $Ni$. The term $E_{many-body}$ arises from indirect interactions, e.g., elastic and $H - H$ interactions, which can be many-body in nature. A CEM is constructed for $E_{many-body}$. A CEM for $NiH_x$ system has not been reported previously.

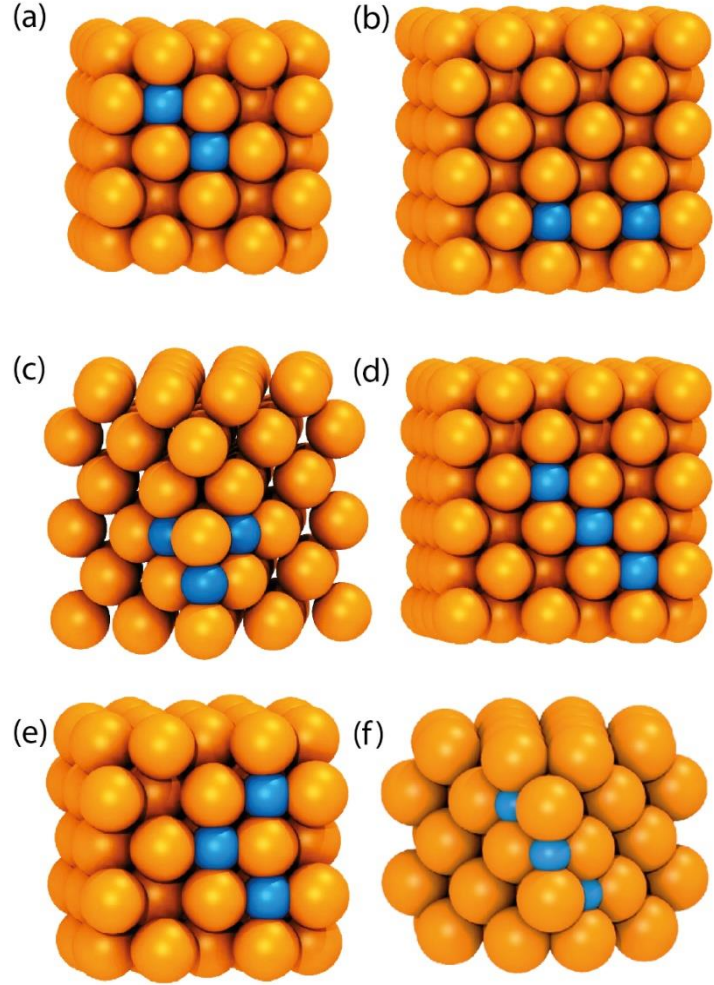

Figure 4. Illustration of clusters included in the CEM. (a) 1NN pair, (b) 2NN pair, (c) T-triplet, (d) I-triplet, (e) L-triplet and (f) V-triplet (see Supporting Information for internal coordinates).

The CEM is a linear expansion containing contributions from discrete clusters:

$$E_{many-body}(x,z) = \sum_c w_c(x) n_c(x,z). \qquad (14)$$

Here $c$ is a type of cluster (see Figure 4), $n_c$ denotes the number of clusters of that type, and $w_c$ is the effective cluster interaction (ECI). Combining Equations (13) and (14):

$$E_{binding}(x,z) = E^0_{binding}(x) + \frac{1}{N_H}\sum_c w_c(x)n_c(x,z) \qquad (15)$$

The parameters $E^0_{binding}$ and $w_c$ need to be determined. Finally, the energy of any configuration is obtained as

$$E_{NiH_x}(x,z) = E_{Ni} + N_H E_H + N_H E^0_{binding}(x) + \sum_c w_c(x)n_c(x,z) \qquad (16)$$

Equation (16) is used for growth rate calculations and calculation of the chemical potential. The expansion in Equation (14) is exact if all possible clusters are considered but in practice it is truncated to include the short-range clusters. The clusters included in the CEM, namely, pairs (first, second, third and fourth neighbor) and triangular (T-), I-, L- and V-triplets, are shown in Figure 4 (see Supporting Information for relative coordinates). The corresponding $w_c$ for these eight cluster types are determined by fitting to the model to a training dataset comprising of different configurations. Put together, a total 1400 configurations are generated for 16 hydrogen concentrations in $0.001 \leq x \leq 0.999$. To capture the variability in $\{n_c\}$ many of these configurations are random realizations of the same $(x,z)$.

Each of the 16 hydrogen concentrations is considered separately, i.e., 16 different CEMs are constructed, as they are more successful in capturing the effect of the short-range ordering and the lattice constant as a function of $x$. For comparison, when one CEM was trained for the entire $x$-space, such a model was found to be less accurate. For each $x$, a dataset typically contains 50-100 configurations (see Section S2 of Supporting Information). The dataset is further partitioned into training and test dataset in a 70:30 ratio. The training dataset provides the fitted $w_c$. Overfitting is avoided by comparing the fit to a test dataset.

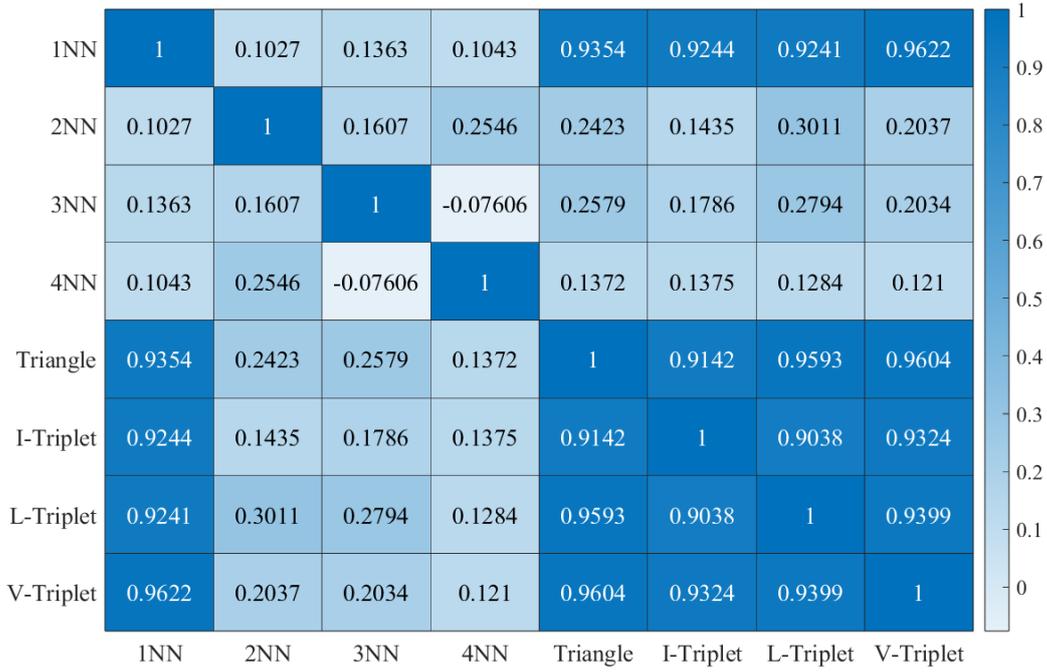

Figure 5. Correlation between different cluster counts for hydrogen concentration $x = 0.2$.

The cluster counts $n_c$ in Equation (14) can be correlated. To give an example, the number of $H - H - H$ triplets is correlated to the number of $H - H$ pairs. Figure 5 shows the correlation matrix for $x = 0.2$. Correlation between 1NN, 2NN, 3NN and 4NN pair counts is weak, but 1NN clusters are highly correlated with the triplets. In conventional regression analysis, it is strictly required that the independent variables should be uncorrelated. Therefore, a principal component regression (PCR) technique, analogous to the approach in Ref. [53], is used to obtain the ECIs and $E_{binding}^0(x)$. In PCR, regression is performed in a low dimensional space $\{\Lambda_p\}$, such that

$$E_{binding} = w_b' + \frac{1}{n}\sum_p \omega_p \Lambda_p. \qquad (17)$$

where $\Lambda_p$ is expressed as a linear combination of cluster counts

$$\Lambda_p = \sum_c \theta_c (n_c - \bar{n}_c). \qquad (18)$$

$\bar{n}_c$ is the mean value of the cluster count. Use of principal components reduces the number of cluster coefficients that need to be determined and helps prevent overfitting. Combining Equations (17) and (18) can provide the values for $E_{binding}^0$ and $w_c$ in Equation (15).

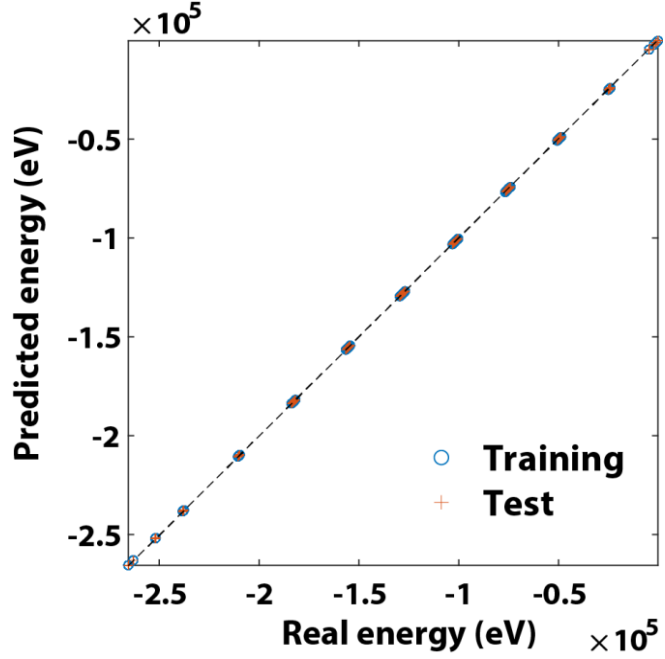

Figure 6. Parity plot for the binding energy calculated from energy minimized configurations (real energy) and the binding energy predicted using the fitted CEM.

Figure 6 shows a parity plot for the real and predicted energies, which confirms that the CEM is accurate. After training the model, the average error in the predicted binding energy per $H$ atom is measured as the normalized root mean squared error (NRMSE)

$$NRMSE = \frac{1}{N_H} \left( \sum_{i=1}^{N_{data}} \frac{\left(E_{many-body,i}^{predicted} - E_{many-body,i}^{real}\right)^2}{N_{data}} \right)^{0.5}. \qquad (19)$$

The NRMSE varies between $0.0026 - 0.5$ meV/H atom (see Supporting Information) and is four orders of magnitude lower than the binding energy per $H$ atom.

Table 1. $E^0_{binding}$ and $w_c$ (in eV) at different hydrogen concentrations.

| x | $E^0_{binding}$ | $w_c$ | | | | | | | |
|---|---|---|---|---|---|---|---|---|---|
| | | 1nn | 2nn | 3nn | 4nn | T-triplet | I-triplet | L-triplet | V-triplet |
| 0.001 | -2.19213 | -0.0277 | -0.0005 | -0.0056 | -0.002 | 0.0003 | -0.0009 | 0.0005 | -0.0021 |
| 0.005 | -2.19265 | -0.0282 | -0.0076 | -0.006 | -0.0011 | -0.0029 | -0.0001 | -0.0016 | -0.0029 |
| 0.01 | -2.19392 | -0.027 | -0.0007 | -0.0031 | -0.0011 | -0.0034 | -0.0011 | -0.0025 | -0.0038 |
| 0.02 | -2.19655 | -0.0263 | -0.0002 | -0.0006 | -0.0002 | -0.0029 | -0.0012 | -0.0025 | -0.0048 |
| 0.1 | -2.22759 | -0.0018 | -0.0007 | -0.0024 | -0.001 | -0.0021 | -0.0012 | -0.0029 | -0.0053 |
| 0.2 | -2.2547 | -0.0013 | -0.0005 | -0.0017 | -0.0007 | -0.0018 | -0.0011 | -0.0024 | -0.0046 |
| 0.3 | -2.27881 | -0.0011 | -0.0003 | -0.0012 | -0.0005 | -0.0016 | -0.001 | -0.0021 | -0.0041 |
| 0.4 | -2.30476 | -0.0007 | -0.0002 | -0.0009 | -0.0004 | -0.0013 | -0.0008 | -0.0017 | -0.0033 |
| 0.5 | -2.32747 | -0.0006 | -0.0002 | -0.0006 | -0.0003 | -0.0011 | -0.0007 | -0.0014 | -0.0029 |
| 0.6 | -2.34584 | -0.0004 | -0.0002 | -0.0006 | -0.0003 | -0.0009 | -0.0006 | -0.0013 | -0.0025 |
| 0.7 | -2.36271 | -0.0003 | -0.0001 | -0.0005 | -0.0002 | -0.0008 | -0.0005 | -0.0012 | -0.0022 |
| 0.8 | -2.38024 | -0.0003 | -0.0001 | -0.0004 | -0.0002 | -0.0007 | -0.0005 | -0.001 | -0.0019 |
| 0.9 | -2.39331 | -0.0002 | -0.0001 | -0.0003 | -0.0001 | -0.0006 | -0.0004 | -0.0009 | -0.0017 |
| 0.95 | -2.39926 | -0.0002 | -0.0001 | -0.0003 | -0.0001 | -0.0006 | -0.0004 | -0.0009 | -0.0017 |
| 0.99 | -2.40441 | -0.0002 | -0.0001 | -0.0003 | -0.0001 | -0.0006 | -0.0004 | -0.0008 | -0.0016 |
| 0.999 | -2.41699 | -0.0002 | -0.0001 | -0.0002 | -0.0001 | -0.0005 | -0.0003 | -0.0006 | -0.0012 |

The fitted $w_c$ are listed in the Table 1. Figure 3b shows $E^0_{binding}(x)$ as a dashed line. $E^0_{binding}$ term dictates the overall binding energy. Negative values of $w_c$ indicate attractive $H - H$ interactions at all concentrations. 1NN pair interactions are strong at low values of $x$. The contribution from the 1NN pair to the overall binding energy at $x = 0.1$, which is given by $c_{1NN}e_{1NN}/N_H$, is -0.52, -3.63 and -6.81 $meV$/hydrogen atom for $z = 0.05, 0.35$ and $0.65$, respectively (see configurations in Figure 2). At $x = 0.9$, the contribution from the 1NN pair to the overall binding energy is -1 and -1.1 $meV/atom$ at $z = 0.9$ and $0.96$, respectively. Both cluster count and ECI strength determine the overall contribution. Thus, all clusters contribute towards making binding of the H atom to Ni stronger. The term $E_{many-body}$ determines the down-shift with respect to $E^0_{binding}$ in Figure 3b.

Once the CEMs have been fitted, the energy of any new $NiH_x$ configuration can be estimated using Equation (16). A linear interpolation scheme is used to obtain the ECIs for an $x$ that lies between two values considered in the Table 1. For $x < 0.001$ and $x > 0.999$, the CEMs created for $x = 0.001$ and $0.999$, respectively, are used.

## 4.2 SRO parameter growth rate

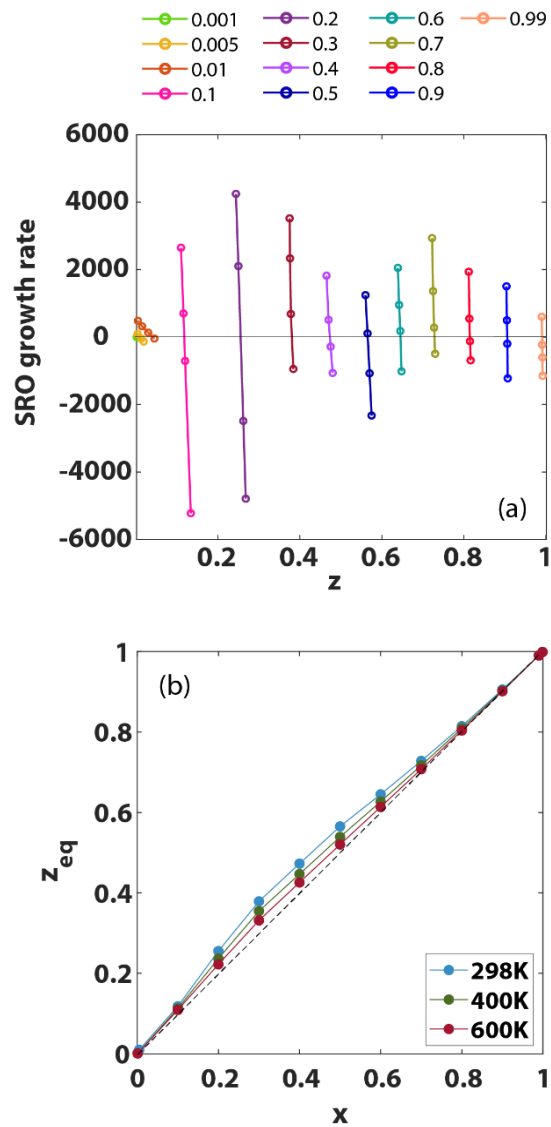

Figure 7. (a) Growth rate for different values of 1NN SRO parameter $z$ ($x$ values are shown in the legend) at $T = 298\ K$. (b) Equilibrium SRO parameter as a function of concentration for different temperatures.

As discussed in Section 2.2, $z_{eq}$ at a given $x$ and $T$ is obtained by identifying the configuration where the growth rate becomes zero. Figure 7a shows several growth rate curves for different values of $x$ (see legend). The growth rate is plotted as a function of $z$. Each open circle corresponds to the growth rate calculated using Equation (5) for a configuration generated with RMC at given $x$ and $z$. The growth rate changes from positive to negative as $z$ increases. For $z < z_{eq}$, the positive growth rate causes the associated configurations to be unstable; $z$ increases. For $z > z_{eq}$, the negative growth rate causes $z$ to decrease. Around $z = z_{eq}$, the growth rate remains small allowing the system to be stable. $z_{eq}$ is obtained by fitting a quadratic equation to the growth rate data (shown in Figure 7a as lines) and identifying the root of the equation. A similar procedure is adopted at other temperatures.

The growth rate for other types of clusters can be calculated in a manner analogous to Equation (5). $v_{i \to j}$ in Equation (5) now pertains to the new cluster count in question. The growth rate for all clusters is found to be nearly zero (not shown) at the calculated $z_{eq}(x, \{w_c\}, T)$. This has some important implications. First, it implies that all cluster counts and the local configuration ceases to evolve, as expected from an equilibrium configuration [41,52]. Second, $z$ is a descriptor for the local material structure. Seven other SRO parameters, including the 2NN, 3NN and 4NN, which comprise of peaks in the radial distribution function, are dictated by $z$. Thus, by specifying $z$ one constrains the allowed probabilities for the additional SRO parameters to a very narrow range. The configuration at $z_{eq}$ is in equilibrium.

$z_{eq}$ is plotted versus $x$ at three temperatures in Figure 7b. At low ($x < 0.02$) and high concentrations ($x > 0.98$), we find that $z_{eq} \approx x$, which implies $H$ and $V$ are perfectly well-mixed. The pair and triplet counts can be estimated using the mean-field approximation. For intermediate concentrations, $z_{eq} > x$ suggesting that the H atoms cluster and that the mean-field approximation is no longer valid. At higher temperatures, $z_{eq}$ approaches $x$, i.e., mean-field approximation is more reasonable here.

## 4.3 Pressure-composition-temperature (PCT) isotherm and phase diagram

Equation (6) is used to calculate the absorption isotherm, i.e., chemical potential $\mu_H$ as a function of $x$ and $T$. The effect of many-body interactions and vibrational corrections on the absorption isotherm is presented here. If $\Delta U_{ins} = 0$, the excess term in Equation (7) becomes zero, which is equivalent to the Langmuir isotherm. Expanding $\Delta U_{ins}$ in terms of the interactions yields

$$\mu(x,T) = k_B T \ln \frac{x}{1-x} + E_{binding}^0(x) \tag{20}$$

$$- k_B T \ln \langle \exp\left(-\beta \sum_c e_c \Delta n_c\right) \rangle_x.$$

The first term is the ideal part, the second term results from the $H - Ni$ interactions and the third term is the configurational part. The $E_{binding}^0$ term causes a significant negative shift in the chemical potential. Figure 8a shows the shifted Langmuir isotherm (green line) at 298 K. An S-shaped curve, which is indicative of hysteresis, is seen. The S-shape arises because the $E_{binding}^0(x)$ is not constant and varies approximately linearly in Figure 3b, analogous to the Temkin isotherm [58]. The attractive many-body term results in an additional negative shift (dashed blue line). Finally, incorporating the vibrational terms results in a positive shift to the isotherm (solid blue line). The vibrational chemical potential $\mu_{vib}$ is found to be constant (0.041 eV) over the entire $x$-range. The experimental absorption data (circles) is close to our simulation results. Figure 8a provides validation of our approach. The hydrogen gas pressure $p_{H_2}$ is related to the chemical potential through Equation (9), which yields the pressure-composition isotherm (see second x-axis). Figure 8b shows a comparison of the isotherms with (solid lines) and without (dashed lines) vibrational corrections. $\mu_{vib}$ is 0.044 eV and 0.064 eV at 400 and 600 K, respectively. The main observation is that the hysteresis width decreases at higher temperatures.

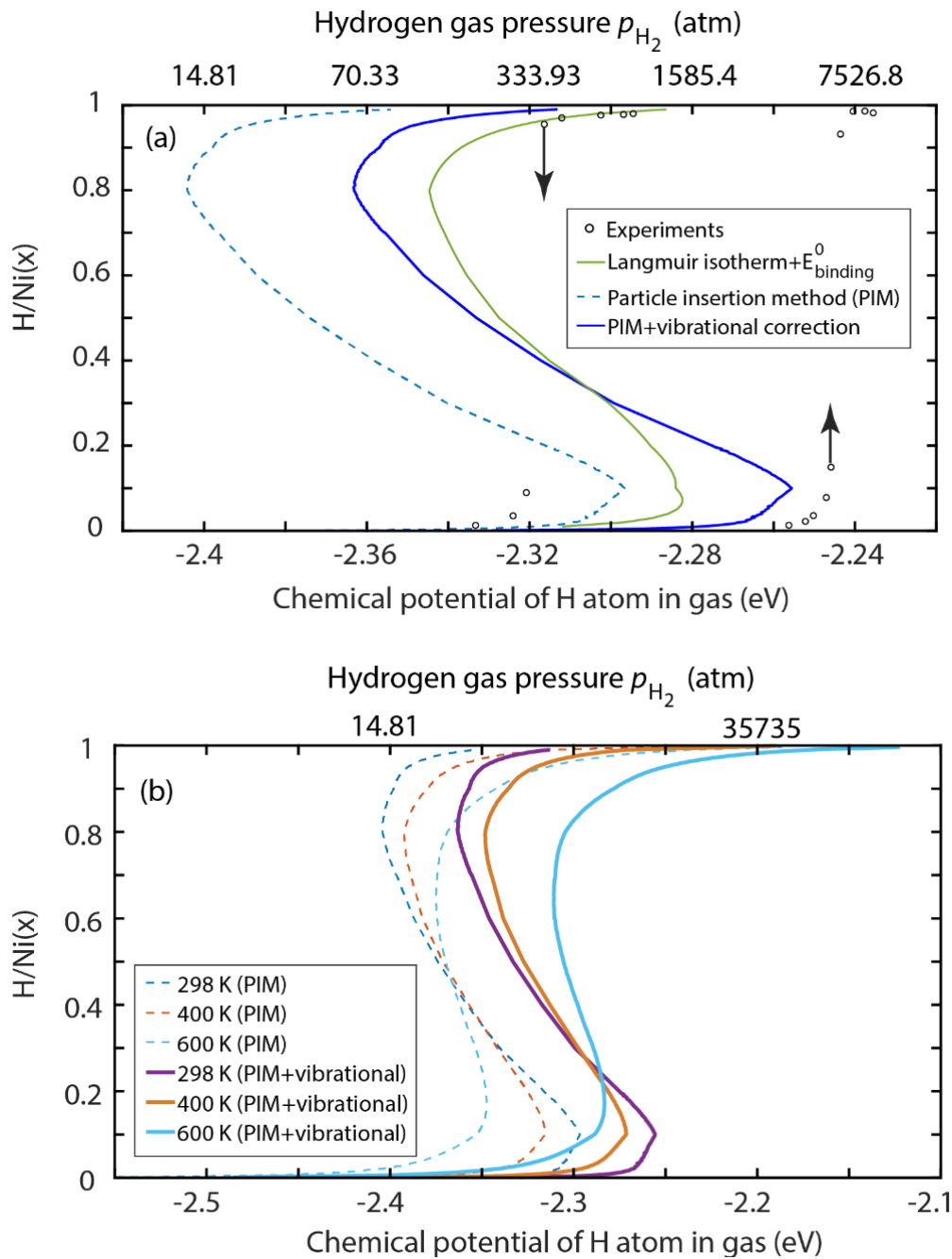

Figure 8. (a) Hydrogen chemical potential in bulk Ni at 298 K. Three levels of approximations are employed. (b) Pressure-composition-temperature (PCT) isotherm with and without vibrational corrections.

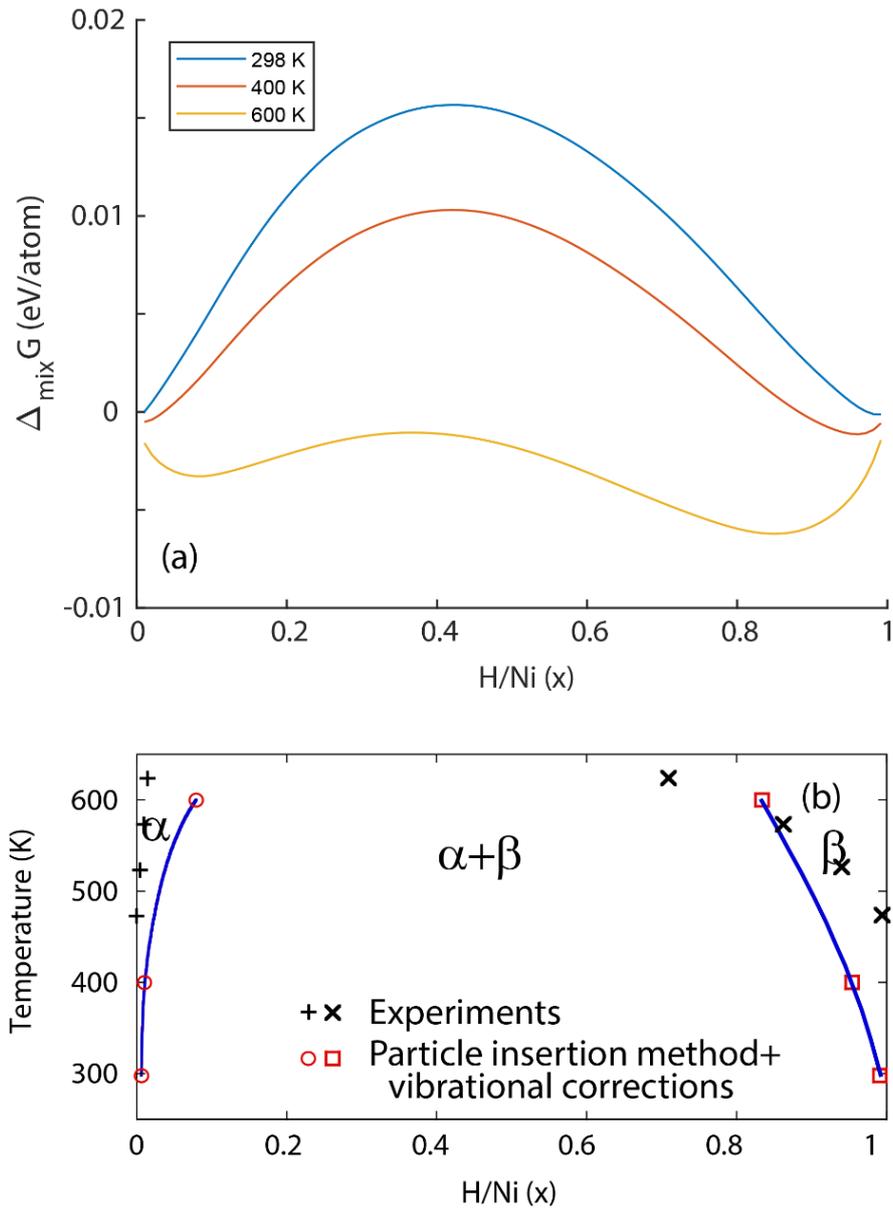

Figure 9. (a) Free energy of mixing as a function of the hydrogen concentration. Vibrational corrections are included. (b) Corresponding phase diagram using the common tangent method. Experimental data is obtained from Wayman et. al [45].

The procedure for construction of phase diagram from the absorption isotherm can be found in any standard thermodynamics textbook [59,60]. Figure 9a shows the Gibbs free energy of mixing at the three temperatures. The curve contains two minima indicating a bistable system.

Two phases ($\alpha$ and $\beta$) are said to be in coexistence where a line acts as a common tangent at concentrations near the left and right minimum. These concentrations are plotted in Figure 9b to obtain the phase diagram. The left (open circles) and right (open squares) curves represent the H concentration in the $\alpha$ and $\beta$ phases, respectively, where the two phases can coexist. The average behavior of the experimental phase diagram data is shown for comparison. In the original experimental reference (see Figure 5 of Ref. [45]) the datapoints for the $\beta$-phase is scattered with the spread in $x$ being 0.1-0.2 for any given temperature ($\beta$-phase lies in the range $0.8 < \beta_{min} < 1.0$). The cross symbols in Figure 9b are therefore an average of the experimental data. The $\alpha$-phase shown in plus symbols denotes the terminal solubility of hydrogen in Ni with $0.02 < \alpha_{max} < 0.1$. Unlike the $\beta$-phase, the reported $\alpha$-phase concentrations are continuous. Typically, long experimental timescales are required with bulk metals to reach equilibrium. This together with the high hydrogen pressures required for achieving the $\alpha$- and $\beta$-phase formation makes experimental observations fairly challenging. The need for computational methods for analysis of phase behavior of such systems becomes imperative.

## 5. Conclusions

The calculation of thermodynamic phase diagram of bulk nickel hydride using reverse Monte Carlo (RMC) has been demonstrated. Analogous to the traditional application of the RMC method, we are able to generate 3D reconstructions of atomic configurations by specifying short-range order (SRO) parameters. The key development is being able to identify which of these configurations corresponds to the equilibrium. This step is achieved by calculating the growth rate for each configuration and determining the equilibrium SRO parameter where the growth rate becomes zero.

To speed-up the growth rate calculations, we constructed a cluster expansion model (CEM) for the hydrogen insertion in Ni by calculating the binding energy of hydrogen using RMC configurations and then estimating the cluster interactions through principal component regression. The CEM confirms that the hydrogen arrangement in the lattice is the key configurational aspect which is captured through the equilibrium SRO parameter at distinct concentrations. The isotherms obtained through the present calculations compare favorably with the experiments. The phase-diagram for the nickel-hydrogen system is generated. The approach is found to be more viable in comparison to the GCMC calculations in terms of computation time. We believe that such a workflow can be in general beneficial for rapidly calculating thermodynamics properties while addressing the combinatorial effect of pressure, composition, temperature, nanostructuring, strain, dopants and impurities, etc. This work not only serves as a template for other studies on hydride materials, but also for oxide, sulfide and many others involving species absorption in a host material.